\documentclass{aa} 
\usepackage{amsmath}
\usepackage{graphicx}
\usepackage{txfonts}
\usepackage{natbib}
\usepackage{enumerate}
\usepackage{amssymb}

\bibpunct{(}{)}{;}{a}{}{,} 

\newcommand{\gr}[1][]{$\gamma$-ray#1}
\newcommand{\pB}{\ensuremath{\varepsilon_\mathrm{B}}}
\newcommand{\ph}{\varepsilon}
\newcommand{\IC}{\mathrm{IC}}
\newcommand{\PP}{\mathrm{PP}}
\newcommand{\nG}{\,\mathrm{nG}}
\newcommand{\es}{1ES~1101-232}

\begin{document}
\title{Cascading on extragalactic background light}

\author{P. d'Avezac \inst{1}
\and G. Dubus \inst{1,2,3}
\and B. Giebels \inst{1}}
\institute{Laboratoire Leprince-Ringuet, UMR 7638 CNRS, Ecole
  Polytechnique, 91128 
Palaiseau, France \and Institut d'Astrophysique de Paris, UMR 7095
CNRS, Universit\'e Pierre \& Marie Curie, 98 bis bd Arago,
75014 Paris, France \and  Laboratoire d'Astrophysique de Grenoble, UMR
5571 CNRS, Universit\'e Joseph Fourier, BP 53, 38041 Grenoble,
France}

\date{Received; accepted}

\abstract
{High-energy \gr[s] propagating in the intergalactic medium can
interact with background infrared photons to produce $e^+ e^-$ pairs,
resulting in the absorption of the intrinsic \gr\ spectrum. TeV
observations of the distant blazar \object{1ES 1101-232} were thus
recently used to put an upper limit on the infrared extragalactic
background light density.}
{The created pairs can upscatter background photons to high energies,
which in turn may pair produce, thereby initiating a cascade. The
pairs diffuse on the extragalactic magnetic field (EMF) and cascade
emission has been suggested as a means for measuring its
intensity. Limits on the IR background and EMF are reconsidered taking
into account cascade emissions.}
{The cascade equations are solved numerically. Assuming a power-law
intrinsic spectrum, the observed 100 MeV - 100 TeV spectrum is found
as a function of the intrinsic spectral index and the intensity of the
EMF.}
{Cascades emit mainly at or below $100\,\mathrm{GeV}$. The observed
TeV spectrum appears softer than for pure absorption when cascade
emission is taken into account. The upper limit on the IR photon
background is found to be robust.  Inversely, the intrinsic spectra
needed to fit the TeV data are uncomfortably hard when cascade
emission makes a significant contribution to the observed
spectrum. An EMF intensity around $10^{-8}$~nG leads to a
characteristic spectral hump in the GLAST band. Higher EMF intensities
divert the pairs away from the line-of-sight and the cascade
contribution to the spectrum becomes negligible.}
{}
\keywords{Radiation mechanisms: non-thermal -- BL Lacertae objects: individual: 1ES 1101-232 -- intergalactic medium -- diffuse radiation -- Gamma rays: observations}

\maketitle

\section{Introduction} 
The observed very high energy (VHE) spectra of extragalactic sources
are attenuated by pair production (PP) on background photon fields. At
energies $\epsilon=100\,\mathrm {TeV}$, interactions with CMB photons
make the universe opaque beyond a few Mpc \citep {GouldSchreder1966}.
In the 1-10 TeV range, the target photon field is the infrared
extragalactic background light (EBL), with an undetermined horizon
$z\ga 0.2$ due to uncertainties in the EBL density at optical to UV
wavelengths. Conversely, observations of absorbed VHE spectra of
blazars can constrain the EBL density at these wavelengths, provided
their intrinsic emission is known
\citep{SteckerSalamonJager1992,Biller1995}.  Recently, the HESS
collaboration used its observations of \es\ ($z=0.186$), together with
a reasonable assumption on the intrinsic spectrum, to estimate a
stringent upper limit to the EBL~\citep{IRHESS}. This estimation did
not consider emission from the cascade initiated when the created
pairs upscatter EBL photons back to VHE energies \citep
{Protheroe1986,Protheroe1993,Aharonian1994,Biller1995,Aharonian2002}.
Cascade emission may make the universe appear more transparent than
under the assumption of pure absorption. Inversely, including
  cascade emission when deconvolving for propagation effects on a
  given EBL leads to intrinsic spectra different from the pure
  absorption case.

The impact of this emission on the EBL upper limit, as derived from
the \es\ spectrum, is considered here. Electrons may diffuse on the
extragalactic magnetic field (EMF), causing their emissions to be lost
for the observer~\citep {Protheroe1986,Aharonian1994,Plaga1995}. The
EBL and EMF governing the propagation of the cascade are described in
\S2. The cascade equations and numerical method are described in
\S\ref{sec:Equations} and applied to the case of \es\ in \S4. The
implications on the EBL and EMF limits are set out in \S5.

\section{Extragalactic backgrounds}
\label{sec:Players}
\begin{figure}
  \resizebox{\hsize}{!}{\includegraphics{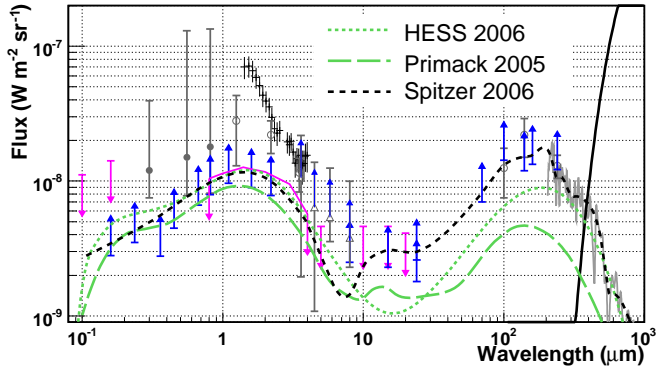}}
  \caption{EBL+CMB photon density (at $z$=0) used in this work. The
  {\em HESS 2006} and {\em Primack 2005} EBL derive from a simulation
  of galaxy formation
  \citep[respectively]{Primack1999,Primack2005}. The {\em Spitzer
  2006} EBL is a best fit to available observations \citep[][ from
  which the measurements shown here were also taken]{Spitzer2006}.}
  \label{fig:EBL}
\end{figure}

\begin{figure}
  \resizebox{\hsize}{!}{\includegraphics{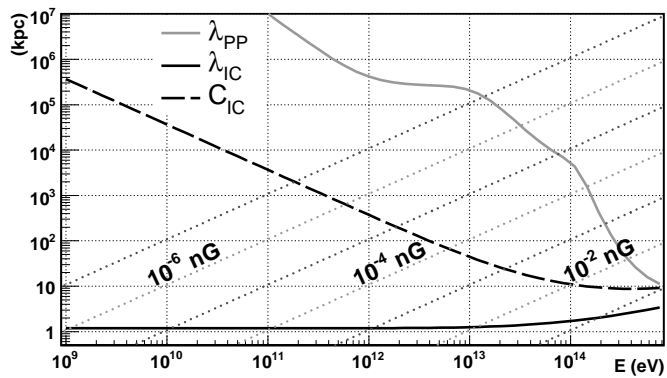}}
  \caption{Mean free path $\lambda_\PP$ for PP on the EBL+CMB
  background as a function of the VHE photon energy. The mean free
  path $\lambda_\IC$ and IC cooling length $C_\IC $ of the pairs on
  the CMB is also shown as function of electron energy. $C_\IC$
  flattens (then rises) at high energies as the interaction enters the
  Klein-Nishina regime. The electron then loses its energy in a single
  interaction ($\lambda_\IC \approx C_\IC$). IC losses of the pairs
  take place on a small scale compared to the $\gamma$-ray attenuation
  length for photon energies $< 300$ TeV. The electron gyroradius
  $R_\mathrm{L}$ for various EMF intensities is indicated by dashed
  diagonal lines. The pairs are expected to be isotropised by the EMF
  for energies and B intensities such that $R_\mathrm{L}\leq C_\IC$. }
  \label{fig:ICLambda}
\end{figure}

In addition to the CMB, the diffuse photon background is constituted
of integrated emission from stars (peaking around $2\,\mathrm{\mu m}$,
see Fig.~\ref{fig:EBL}) and heated dust (peaking around
$200\,\mathrm{\mu m}$). The EBL spectral energy distribution is
difficult to measure directly because of the strong zodiacal and
galactic foregrounds. Lower limits have been set using source counts
while SED shapes have been derived from simulations of galaxy
formation \citep{PrimackReview,Lagache2003,Xu2001}. The EBL shape {\em
HESS 2006} (Fig.~\ref{fig:EBL}) was computed by \citet{Primack1999}
and normalised by a factor 0.45 in \citet{IRHESS} to account for the
TeV observations of \es. {\em Primack 2005 } refers to an EBL
normalised instead to fit the lower limit set by galaxy counts. The
{\em Spitzer} observations suggest higher fluxes in the
$10-1000\,\mu\mathrm{m}$ range~\citep {Spitzer2006}. This affects
attenuation above 20 TeV but has been verified to have no consequence
on the results presented here.

The created pairs can be deflected from the line-of-sight by an
extragalactic magnetic field (EMF).  Faraday rotation and synchrotron
emission in radio yield estimates of magnetic fields in galaxies
(roughly $>10\nG$), or in clusters ($\leq 0.1-1\nG$) and even some
super- clusters
($\leq\nG$)~\citep{Kronberg1994,Widrow2002,Vallee2004}. The EMF
outside these structures is unconstrained and may be as low as
$10^{-19}\nG$ \citep[and references therein]{Fan2003}.  For such very
weak EMFs, the deflection of electrons due to IC interactions is
negligible and the cascade occurs along the line-of-sight with a short
delay of the secondary emission
\citep{Plaga1995,Cheng,Dai2002}. Diffusion on a stronger EMF
creates a halo around \gr\ sources and isotropizes the cascade
emission \citep {Aharonian1994}.  This occurs when the gyroradius
$R_\mathrm{L}$ of the pairs is much lower than their Compton cooling
length $C_\IC=E (dE/dl)^{-1}_\IC$. Since mostly CMB photons are
upscattered, the minimum $B$ required to isotropise pairs of energy
$E$ is $3\, 10^{-6} E_\mathrm{TeV}^2 (1+z)^4\nG$. Much of the
isotropic re-emission is lost to the observer and the pairs diffuse on
a scale $\sim (R_\mathrm{L} C_\IC)^{1/2}$. For intermediate EMFs, the
TeV electrons in the beamed relativistic jet are deflected by $\sim
{C_\IC}/{R_\mathrm{L}}$. Halo sizes $\ga 0.1^\circ$ could be resolved
by \gr\ detectors and used to estimate the EMF
intensity~\citep{Neronov2006}. Photons in 0.1$^\circ$ haloes have
propagation times varying by $\sim 10^5$ years, averaging out any time
variability \citep{Fan2003}.  In the following, the cascade emission
is assumed to be unresolved from the source and delays are not
considered. The TeV emission detected by HESS from \es\ appears
  to be at a low flux level with no significant variability.

\section{Cascade equations}\label{sec:Equations}

The cascade is described by a set of two coupled equations involving
the photon energy density $n_P(\ph)$ and the electron (positron)
energy density $n_E(E)$: \begin{align} \label{eq:PropPP}
  c\partial_t n_P= -\frac{1}{\lambda_\PP} n_P
     &+c_B\int^{+\infty}_{\ph}\mathrm{G_{IC}}(e,\ph)\,n_E(e)\mathrm{d}e\\
  \begin{split}
  c\partial_t n_E= -\frac{1}{\lambda_\IC}n_E
     &+2\int^{+\infty}_{E}\mathrm{G_{PP}}(e, E)\, n_\ph(e)\mathrm{d}e\\
     &+\int^{+\infty}_{E}\mathrm{G_{IC}}(e, e-E)\, n_E(e)\mathrm{d}e
  \label{eq:PropIC}
  \end{split}
\end{align}
The first term in both equations is the sink term due to PP
(Eq.~\ref{eq:PropPP}) or IC losses
(Eq.~\ref{eq:PropIC}). $\lambda_\PP$ and $\lambda_\IC$ are the mean
free path for each interaction. The second term is the source term
corresponding to cascade emission (Eq.~\ref {eq:PropPP}) or pair
creation (Eq.~\ref{eq:PropIC}, with a factor 2 for the pair). The
cascade emission factor $c_B$ is $1$ when the EMF is ignored, and
approximated to 0 when the electron population is considered
isotropised. The pair production term is written in terms of
$\mathrm{G}_\PP(\ph,E)= \int \partial_E\sigma_\PP(\ph, \pB)
u(\pB)d\pB$, where $ \partial_E\sigma_\PP$ is the differential
cross-section and $u$ is the photon background energy density
(EBL+CMB). The IC radiation term $\mathrm{G}_\IC(E,\ph)$ is defined
similarly. The third term in Eq.~\ref{eq:PropIC} reflects IC cooling
of electrons from higher energies. All of these terms are functions of
$z$.

The integrated cross-sections for PP and IC on isotropic target
photons are taken from \citet {GouldSchreder1966} and
\citet{Jones1967}. Analytic expressions of the differential cross-
sections derived by \citet{Zdziarski1988} for background densities in
the form of blackbodies or power laws are used to calculate
$\mathrm{G}_\PP$ and $\mathrm{G}_\IC $. The cascade equations are
solved numerically by combining $n_P$ and $n_E$ into a single vector
$V$ defined on a logarithmic scale of energies
$(\varepsilon_0\zeta^i)$, from $\varepsilon_0$=10$^7$ eV up to
10$^{17}$ eV in 250 steps (Thus $\zeta= (10^{17}/
10^{7})^{1/250}$). To ensure energy conservation, the integrals on
$\mathrm{G}_\PP$ and $\mathrm{G}_\IC$ are calculated as
\begin{equation}
    \int \mathrm{G_{IC}}(e,\ph)\,n_E(e)\mathrm{d}e=
    \sum_k
        V_{k,E}\int_{\zeta^{-1/2}}^{\zeta^{1/2}} 
        \frac{\epsilon_0\zeta^k u\mathrm{G_{IC}}(\epsilon_0\zeta^k\,u,
\ph)\mathrm{d}u}{\zeta^{1/2}-\zeta^{-1/2}}
\end{equation}
The cascade equations may then be rewritten as a matrix $\mathrm{\bf
P}$ acting on the vector $V$ : $V(t+\delta t)= \exp(\delta
t\mathrm{\bf P}) V(t)$ ($\exp$ is developed to the $4^{th}$ order in
$\delta t$). The terms in $\mathrm{\bf P}$ are of the order of $
\lambda_\IC^{-1}$ or less, hence it is enough to take steps of size
$c\delta t= 0.1\, \mathrm{kpc}$, updating the matrix $\mathrm{\bf
P}(z)$ every $\delta z= 0.001$ with $ \mathrm{d}z= {H_0\,(1+z)}
[\Omega_M\,(1+z)^3+\Omega_\Lambda
+(1-\Omega_M-\Omega_\Lambda)\,(1+z)^2]^{1/2} \,\mathrm{d}t$ and values
for $H_0 $, $\Omega_M$ and $\Omega_{\Lambda}$ taken from
WMAP~\citep{WMAP2003}. Thus, at $z= 0.2$, $\delta z\approx 3\,10^4
c\delta t$.

\section{Application to \es}
\label{sec:Results}
\begin{figure}
  \resizebox{\hsize}{!}{\includegraphics{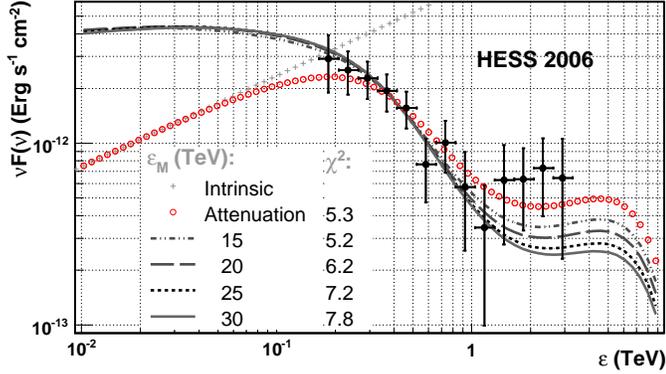}}
  \caption{\es\ observed and modelled spectra with a maximal ({\em
  HESS 20006}) EBL and including cascades with no magnetic field. HESS
  observations points are in black \citep {IRHESS}. Markers indicate
  the attenuation only observed spectrum (circles) and the
  corresponding intrinsic spectrum (crosses), whereas the lines
  indicate the observed spectra with cascade emissions. Intrinsic
  spectra are in the form of $\nu F_\nu \propto E^{0.5}
  {\mathrm{d}\nu}$ and adjusted to the data.  Cascade emission
  accumulates at $100\, \mathrm{GeV}$ and below, softening the spectra
  compared to pure absorption. The HESS upper limit on the EBL remains
  valid after taking the full emission from cascades into account.}
\label{fig:EmaxHigh}
\end{figure}

\begin{figure}
  \resizebox{\hsize}{!}{\includegraphics{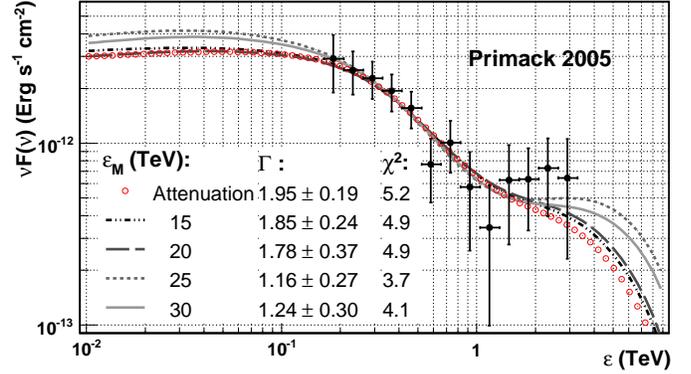}}
  \caption{\es\ observed and modelled spectra with a minimal ({\em
  Primack 2005}) EBL and including cascades with no magnetic
  field. The intrinsic spectrum is now adjusted to the HESS data
  leaving the spectral index $\Gamma$ free. For pure absorption, the
  best index is $\Gamma= 1.95$. With cascades, the index hardens as IC
  emission softens the propagated spectrum. For high
  $\epsilon_\mathrm{M}$, the best index softens again so as to limit
  the amount of cascading but the fit worsens. Significant cascading
  on the minimal EBL and in a very weak EMF implies a very hard input
  spectrum in order to account for the observations.}
  \label{fig:EmaxLow}
\end{figure}

The SED of the attenuating EBL can be deconvolved from \gr\
observations of extragalactic sources (TeV blazars), given {\em a
priori} knowledge on the intrinsic spectra. Modelling observed spectra
as power-laws, the effect of PP is to soften the intrinsic spectral
index, increasingly so with EBL intensity.  Hence, using observations
of the farthest TeV blazar and assuming the hardest possible intrinsic
spectrum puts an upper limit on the EBL responsible for
attenuation. Current theoretical understanding of shock acceleration
limits the intrinsic particle distribution in blazars to a power-law
of index no harder than a 1.5 and correspondingly, an intrinsic photon
spectrum $\mathrm{d}N\propto E^{-\Gamma} \mathrm {d}E $ with
$\Gamma\geq$ 1.5 \citep{IRHESS}.

\es, at $z= 0.186$, is currently the farthest known TeV source and was
used by the HESS collaboration to set an upper limit to the EBL
corresponding to the {\em HESS 2006} SED shown in Fig.~1. The
comparison between a $\Gamma$=1.5 power-law attenuated by the {\em
HESS 2006} EBL (without cascade, $c_B$=0) and the data is shown in
Fig.~3, reproducing the results of \citet{IRHESS}. Attenuated spectra
taking into account the full cascade emission with $c_B$=1 ({\em i.e.}
a null EMF) are also shown for various values of the maximum energy
$\epsilon_\mathrm{M}$ to which the intrinsic power-law extends.  Since
cascades initiated at higher energies increase the photon populations
in lower ones, one might expect the final spectra to appear harder
than for pure absorption. However, because IC occurs predominantly on
the CMB, the cascade emission accumulates below $100\,\mathrm{GeV}$,
softening the spectrum between 100~GeV and 1 TeV. High values of
$\epsilon_\mathrm{M}$ lead to more cascading and more softening. The
$\chi^2$ values suggest $\epsilon_\mathrm{M}<15\,\mathrm{TeV}$,
although further observations, particularly above 1 TeV, would be
necessary in order to confirm this. For such low $
\epsilon_\mathrm{M}$ values, not many photons initiate cascades. For
higher $\epsilon_ \mathrm{M}$, the softening is such that a lower EBL
would be needed to match the data.  Thus the {\em HESS 2006} upper
limit found by \citet{IRHESS} holds strong, even in this extreme limit
where all the cascade emission is received by the observer.

Inversely, the intrinsic \gr\ spectrum at the source can be obtained
given some assumption on the intervening EBL. Using the lower limit on
the EBL set by galaxy counts ({\em Primack 2005} in Fig.~1) gives a
limit on how soft the intrinsic spectrum can be. For pure absorption,
the best fit has $\Gamma= 1.95\pm0.19$ (Fig.~4). As expected, this is
softer than the $\Gamma=1.5$ assumed above, yet still suggests that a
good fraction of the \gr\ energy in \es\ is output above a TeV. A
  hard $\Gamma\leq 2$ intrinsic spectrum is needed if cascade emission
  is to contribute significantly to the low-energy continuum
  \citep{Aharonian2002}. \es\ is the first blazar where the intrinsic
  spectrum is constrained to be hard enough for this, even in the
  minimal EBL limit. 

Including cascade emission in the fit (Fig.~4) hardens even more the
intrinsic spectrum as the cutoff $\epsilon_\mathrm{M}$ increases and
cascades contribute more and more to the observed spectrum. For higher
$\epsilon_\mathrm{M}$, the best fit $\Gamma$ increases again to
mitigate the pronounced softening from the strong cascading but the
fit worsens. This also holds for (implausibly) high values of
  $\epsilon_\mathrm{M}>100$~TeV, for which cascade emission
  largely dominates at a few TeV.  The hard intrinsic spectra found
here, assuming the {\em Primack 2005} is indeed the minimum possible
EBL, suggest either that $\epsilon_\mathrm{M}$ is not greater than a
few TeV, so that there is little cascade emission in the TeV range, or
that a large part of the cascade emission is lost due to diffusion on
the EMF.

\begin{figure}
  \resizebox{\hsize}{!}{\includegraphics{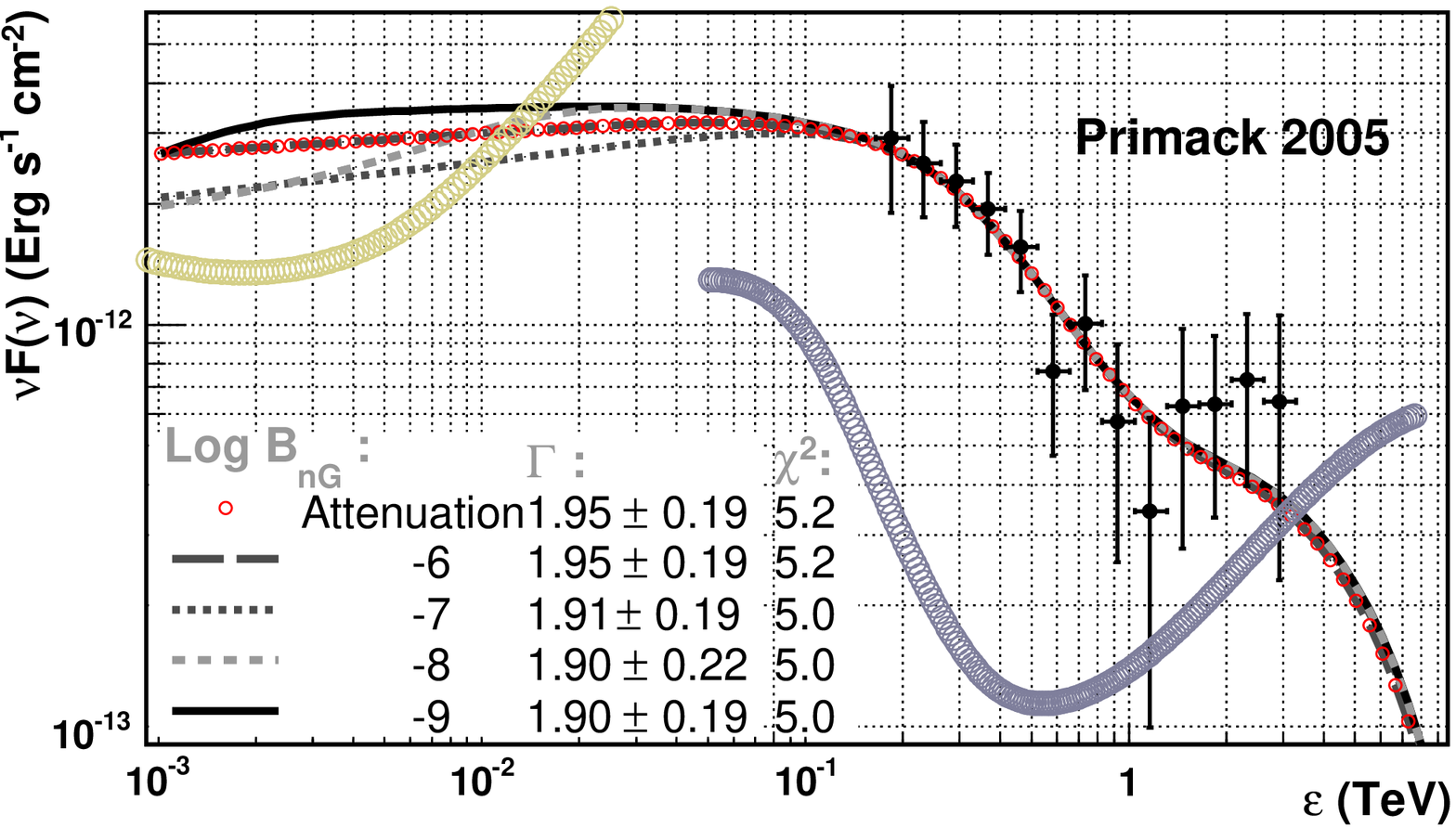}}
  \resizebox{\hsize}{!}{\includegraphics{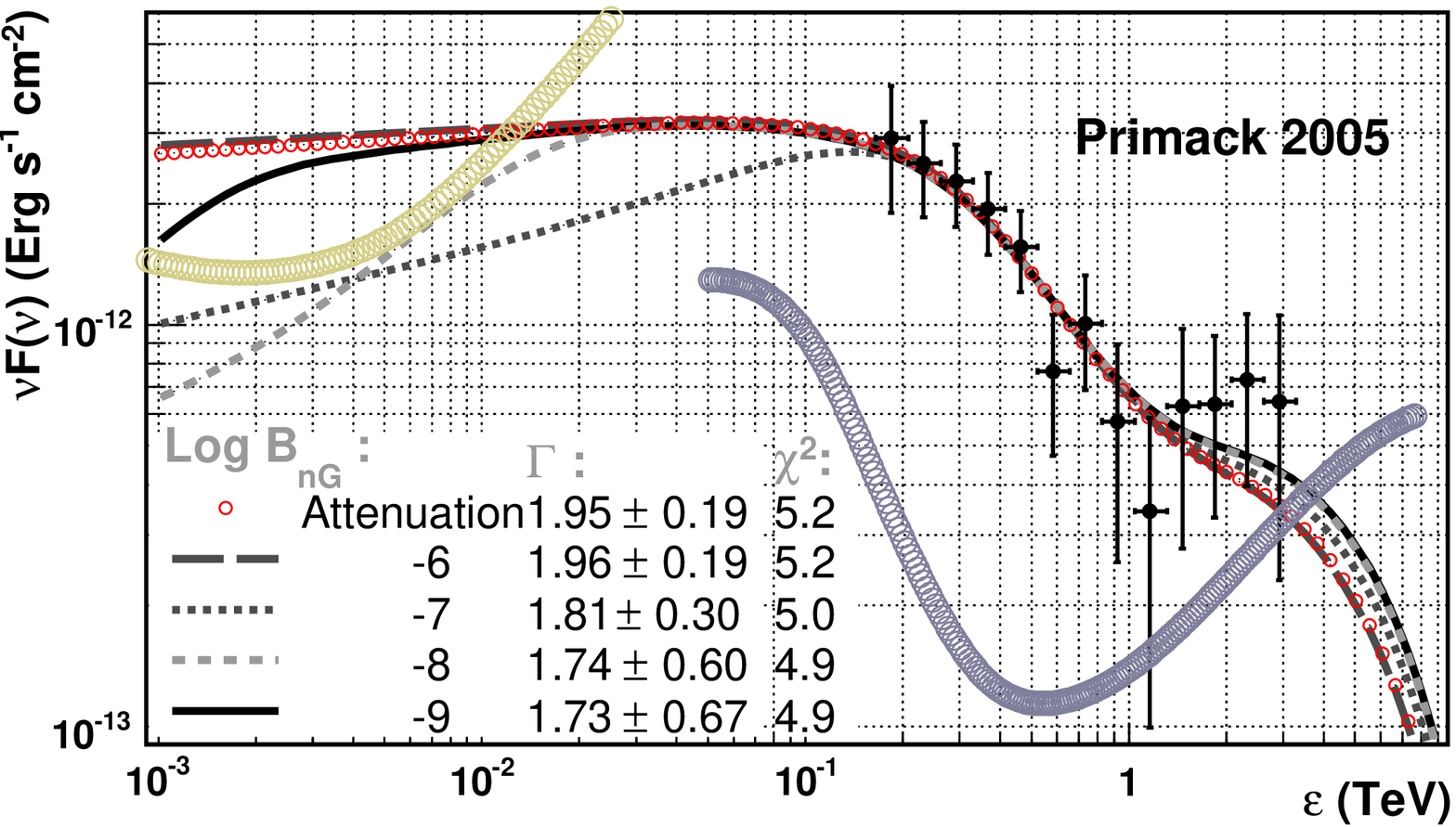}}
  \caption{Observed spectra for the {\em Primack 2005} EBL and various
  EMFs between $10^{-9}$ and $10^{-6}\,\mathrm{nG}$. The spectra are
  adjusted to the HESS points, leaving $\Gamma$ free but fixing
  $\epsilon_\mathrm{M}= 10\,\mathrm{TeV}$ (top) or 20\, TeV
  (bottom). For high EMFs the cascade emission does not reach the
  observer and the spectrum is identical to the pure absorption
  case. For very low EMFs the results are similar to those shown in
  Fig.~4. Intermediate values lead to a more or less pronounced bump
  at 1-100 GeV energies over the intrinsic continuum. Hatches
  represent projected 5-$\sigma$ sensibilities for one year of
  observation with GLAST ($\leq 30$~GeV) and 50 hours with HESS-2
  ($\geq 50$~GeV).}
  \label{fig:EMFLow}
\end{figure}

As discussed in \S2, the electron diffusion on the EMF depends on the
ratio $R_\mathrm{L}/ C_\IC$. The effect on the observed spectra is now
taken into account by setting $c_B$=0 when $R_\mathrm{L}/C_\IC$$<$300
(corresponding to a maximum deviation on the line-of-sight of
  0.1\degr-0.2\degr\ equal to the best GLAST angular resolution) and
$c_B$=1 otherwise. For example, an EMF of $10^ {-6}\,\mathrm{nG}$
means that emission from electrons of energy $E\la20\,\mathrm{TeV} $
is suppressed. This will lead to low-energy cutoff in the cascade
spectrum as only emission from pairs above a certain energy reaches
the observer. The overall spectrum appears as a hump between $\gamma^2
h \nu_{\rm CMB}$ (with $\gamma$ the Lorentz factor of the electrons
for which $R_\mathrm{L}=300 C_\IC$) and 100~GeV (above which
absorption dominates). Hence, a non-zero EMF leads to a reduction of
the overall cascade emission seen by the observer (compared to
Figs.~3-4) but can also lead to a well-defined signature above the
continuum.

Figure \ref{fig:EMFLow} shows the observed spectra for a {\em Primack
2005} EBL and for EMF intensities between $10^{-9}$ and
$10^{-6}\,\mathrm{nG}$. The intrinsic power-law index was left free
but its cutoff $\epsilon_\mathrm{M}$ was fixed at either
$10\,\mathrm{TeV}$ or $20\,\mathrm{TeV}$. The best fit index $\Gamma$
is then found for each value of the EMF. In both cases, the spectra
for an EMF $ \ga 10^{-6}$~nG are not much different from the pure
absorption case as most of the cascade emission is isotropised and
lost to the observer. With $\epsilon_\mathrm{M} $=10~TeV, the best-fit
intrinsic slopes are flat in $\nu F_\nu$ and the cascade emission is
essentially indistinguishable from the GeV continuum for any value of
the EMF. The intrinsic emission is assumed here to be a simple
power-law over the whole energy range. More realistic modelling
would result in a curved intrinsic Compton component. The cascade
emission might then be more readily identifiable over an intrinsic
continuum rising from GeV to TeV energies.

Stronger cascading, as a result of a higher cutoff energy
$\epsilon_\mathrm{M}$ and/or a higher EBL density, makes the hump
apparent for the same reason. The intrinsic spectrum is then
necessarily much harder, enabling the contribution from the cascade to
stand out over the continuum. The bottom panel of Fig.~5 shows that
EMF intensities of 10$^{-9}$--$10^{-8}$~nG can be identified using
GLAST and HESS-2 if $\epsilon_\mathrm{M}$=20~TeV.  Cascade emission is
not diluted for EMF intensities weaker than $10^{-9}$~nG and there is
no spectral feature to measure the EMF. Surprisingly, in most cases
\es\ is only slightly above the GLAST one-year detection limit. Unless
they become active and flaring, low flux state blazars detected by
HESS such as \es\ are likely to be difficult to detect with GLAST,
illustrating the advantage provided by the large collecting area of
ground-based Cherenkov arrays (but at higher energy thresholds).
Similar results are obtained by keeping $\epsilon_\mathrm{M}$ at
10~TeV but using the stronger {\it HESS 2006} EBL. However, in this
case, the fitted intrinsic slopes are very hard ($\Gamma\approx 1.1$)
when the EMF intensities are lower than $10^{-7}$~nG.

The softest values of $\Gamma$, which are the most plausible given the
present knowledge on blazars, favour values of the EMF higher than
$10^{-6}\,\mathrm{nG}$ and/or a cutoff energy below 20~TeV. VHE
emission from nearby, little-attenuated blazars can be investigated
for evidence of cutoffs at energies $>20$~TeV --- although it should
be noted that {\it e.g.} HESS observations of \object{Mkn 421}
($z=0.03$) taken at a high flux actually measure an exponential cutoff
at 3~TeV \citep{2005A&A...437...95A}. EMF intensities $\ga 10^{-6}$~nG
are consistent with measures inside clusters and super-clusters. Such
structures may reach 10--50 Mpc in size, which is greater than the
attenuation length for \gr[s] above $50\,\mathrm{TeV}$.  Furthermore,
the largest voids, where the EMF is expected to be very small, have a
size (20 $h^{-1}$ Mpc, \citealt{Patiri2006}), smaller than the
distance to \es.  Hence, cascades are likely to be initiated inside
walls. As $C_\IC$ is only of the order of $1\,\mathrm{Mpc}$, such
cascades reemit most of their energy within the confines of the
clusters, and thus are subject to diffusion. In this case, the cascade
emission can only be detected by resolving the faint halo surrounding
the \gr\ source.

\section{Conclusion}

The impact of extragalactic cascade emissions on the GeV-TeV spectrum
of \es\ has been investigated and shown to soften the observed
spectrum in the TeV range compared to pure absorption. This occurs
because most of the cascade emissions occurs at $100\,\mathrm{GeV}$
and below. As a result, the upper limits on the EBL determined by HESS
are strengthened in the sense that taking cascades into account would
lead to harder intrinsic spectra than judged plausible, or to a
reduced EBL upper limit. Inversely, using lower limits on the EBL
coming from galaxy counts, the intrinsic spectrum of \es\ is found to
have $\Gamma \leq 1.95$, with very hard values if there is an
important contribution from cascade emission. This is at odds with
current theoretical and observational understanding of blazars. A
cutoff $\la 10$~TeV in the intrinsic spectrum would limit the cascade
contribution. This contribution would also be quenched if the EMF
intensity is greater than $10^{-6}$ nG, as expected away from
  voids. A lower EMF increases the amount of cascade emission
reaching the observer in the GeV band, with a signature in the GLAST
band for intensities $\sim 10^{-8}$~nG --- but at the price of a hard
intrinsic spectrum so as to fit the HESS observations.
 
\bibliographystyle{aa} 
\bibliography{Biblio}

\begin{appendix}
\end{appendix}
\end{document}